\documentstyle[PASJadd,epsf]{PASJ95}
\newcommand{\vol}{00}
\newcommand{\no}{00}
\newcommand{\lett}{}
\newcommand{\spage}{1}
\newcommand{\rdate}{1999 April 5}
\newcommand{\adate}{1999 June 10}

\begin{document}
\setcounter{page}{1}

\title{Gravitational Microlens Mapping of a Quasar Accretion Disk}

\author{Shin {\sc Mineshige} and Atsunori {\sc Yonehara} \thanks{Research Fellow of the Japan Society for the Promotion of Science.} \\
{\it Department of Astronomy, Faculty of Science, Kyoto University, Sakyo-ku, Kyoto 606-8502} \protect\\
{\it  E-mail(SM): minesige@kusastro.kyoto-u.ac.jp}
}

\abst{
%\begin{abstract}
Since the radiation from different portions in the central
region of a quasar
can be successively amplified during a microlensing event,
microlensing light curves provide fruitful information
regarding the emissivity distribution of an accretion disk
located at the quasar center.
We present a basic methodology of how to map the
emissivity distribution of the disk as a function
of the radial distance from the center, $Q(r)$,
from `observed' microlens light curves
during a caustic crossing event.
Our proposed method is 
based on the standard inversion technique, 
the so-called regularization method,
and Abel's transformation of the one-dimensional luminosity profile
integrated along the line parallel to the caustics.
This technique will be used to map the disk structure in
Q2237+0305, for which the HST and AXAF observations are scheduled.
A reconstruction of the image on length scales of several to ten AUs
is quite feasible for this source,
as long as the measuring errors are within 0.02 mag
and the observation time intervals are one week or less.}

\kword{Accretion disks --- Active Galactic Nuclei --- Black Holes --- 
Microlensing --- Quasars: individuals (Q2237+0305)}

\maketitle
\thispagestyle{headings}

\section{Introduction}

In many cases of astrophysical objects,
their direct images are too small to resolve with the usual telescopes.
In the case of a quasar accretion disk, for example,
its angular size is, typically, on the order of
$\theta_{\rm d} \sim 0.01{\rm pc}/1{\rm Gpc} 
                \!=\! 10^{-11} {\rm rad}
                \!\approx\! 1\mu{\rm as}$.
In the case of a binary accretion disk, on the other hand, we estimate
$\theta_{\rm d} \sim 3\times 10^{10}{\rm cm}/1{\rm kpc} 
                \!=\! 10^{-11} {\rm rad}
                \!\approx\! 1\mu{\rm as}$.
These are both considerably below the angular resolution of 
any present-day telescopes.
What is usually attempted is, therefore,
to construct theoretical models based on the basic
equations of (magneto)hydronamics and to infer their structure
from the limited number of information,
such as the total radiation output originating from
their entire surface at some wavelength bands.

In some special cases, however, 
we can `resolve' the spatial structure of the disk
by using a `natural' telescope.
A good example is the technique of eclipse mapping
(Horne 1985).  Since a certain fraction of an accretion disk
is shadowed by a companion star in an eclipsing close binary
and since the shadowed part varies with time
in accordance with the orbital phase,
we can map the disk emissivity distribution from the eclipsing light curves.
A variety of disk luminosity profiles has been revealed with
this technique and our knowledge about the disk structure
has been remarkably enriched.
Unfortunately, however, this technique is irrelevant to probing
a quasar accretion-disk structure 
because there are no eclipsing objects known to date
(see, however, McKernan, Yaqoob 1998).

An alternative and potentially powerful method 
using a microlensing phenomenon has been considered.
Chang and Refsdal (1979, 1984)
considered flux changes that might occur due to gravitational lensing
by a single star in an extended gravitational lens galaxy.
Now, 
a stimulating discussion has started that such a microlensing phenomenon
can be used to investigate the quasar central structure 
(Grieger et al. 1988; Rauch, Blandford 1991) 
as a `gravitational telescope' (Blandford, Hogg 1995).
The Einstein Cross,
Q2237+0305 (e.g., Huchra et al. 1985), is the first object
in which quasar microlensing events were detected 
(Corrigan et al. 1991, Houde, Racine 1994, see also Ostensen et al. 1996).
These observations suggest that microlensing events seem to 
take place almost every year.
This rather high frequency is consistent with the microlens  
optical depth of $\tau \simeq 0.2 - 0.8$ obtained 
by a realistic simulation 
of the lensing galaxy (i.e., Wambsganss, Paczy\'nski 1994).

We, here, specifically consider the microlensing event of this source 
caused by the so-called `caustic crossings.'
Several authors have already calculated and inspired this `caustic' case 
based on simple disk models (e.g., Wambsganss, Paczy\'nski 1991; 
Jaroszy\'nski et al. 1992) or recently on 
the realistic disk models (Yonehara et al. 1998, 1999). 
All of these calculations correspond to the so-called forward problem;
i.e.,
they calculated microlens light curves based on disk models
prescribed {\it a priori}.  In contrast,
we are, in the present study, concerned with a distinct approach 
called the inverse problem; i.e.,
we consider how to reconstruct the disk image (or more precisely, 
the emissivity distribution) of a quasar accretion disk
from microlensing light curves.

For this purpose, we first simply apply the regularization method, 
one of the most well-known non-classical inversion techniques.
The basic methodology of this method was already given 
by Grieger et al. (1991, hereafter GKS).
Indeed, their proposed procedure has been very successful, but 
they have assumed rather idealized situations
and, thus, there remain some problems requiring further investigation
before realistic applications are obtained.
\begin{enumerate}
\item
GKS assumed no radiation from any part outside a caustic.
As a result, they could precisely determine the contribution 
from a small outer portion of the disk to the total light
from the shape of the light curve
at the beginning of a microlens event.  If this were the case,
we would not be able to see the
disk emission in the absence of a microlensing,
contrary to the observations.
\item
The goal of GKS was to reconstruct a one-dimensional luminosity profile 
%as a function of the distance from the line ($\xi$).
[$P(\xi$) in their notation]
which is the luminosity integrated over the disk plane
in the direction parallel to that of a caustic (see figure 1
for the definition of $\xi$).
%%%%%%%%%% Figure 1 %%%%%%%%%%
\begin{figure}[htbp]
 \epsfxsize\columnwidth  \epsfbox{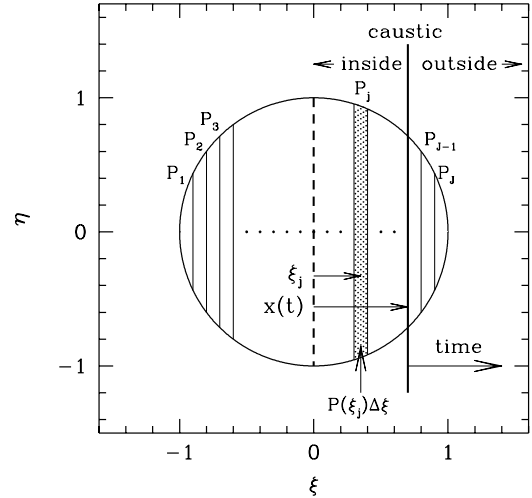}
\caption{Schematic view of a `caustic' crossing during a microlens event.  
The caustic is represented by the thick vertical line,
and the left-hand region corresponds to the parts inside the caustic
and is subject to microlensing light amplification.
One-dimensional luminosity profiles, $P_1, P_2, \cdots, P_J$,
are the integral of the local emissivity of the disk 
along the vertical lines parallel to the caustic.}
\end{figure}
%%%%%%%%%%%%%%%%%%%%%%%%%%%%%%
To make a direct comparison with accretion disk models,
however,
it will be more convenient to express the emissivity distribution
as a function of the distance from the center, $Q(r)$,
since the central part of the disk can be reasonably assumed to
be axisymmetric.
It is thus needed to transform $P(\xi)$ to $Q(r)$.
\item
GKS assumed a rather smooth emissivity distribution around the center,
which certainly makes the analysis easier than otherwise.
According to realistic disk models, however,
it seems more likely that
the emissivity has a power-law dependence on the radius,
%$Q(r) \propto r^{-a}$ with $a$ being a positive parameter,
thus being sharply peaked around the center.
We wish to know what fraction of the total energy output
originates from a compact region in the disk.
\end{enumerate}

The aim of the present study was to improve the GKS method
so that it can be applied to more realistic situations.
Our version of the inverse technique will be described in section 2.
We then present the results of disk mapping
for the calculated microlensing light curve in section 3.
The final section will be devoted to a summary and discussion.

\section{Inversion Procedures}

\subsection{Setting Matrix Forms}

We first describe the regularization method adopted by
GKS with some modifications.

Chang and Refsdal (1979, 1984) have shown
that the flux from the bright double image of a point source 
close to a caustic can be amplified approximately according to
$(x-\xi)^{-1/2}$, where $x$ and $\xi$ are, respectively,
the radial distances of the caustic and that of a part of the disk
in question from a line which is
parallel to the caustic and which crosses the disk center (see figure 1).
Note that $x$ is time-dependent; i.e.,
\begin{equation}
\label{vcaus}
   x = x(t) = V_{\rm caus}t/r_0,
\end{equation}
where the unit of length, $r_0$, is taken to be a typical disk dimension,
$t=0$ corresponds to the time when a caustic crosses the center of the disk,
and $V_{\rm caus}$ is the transverse velocity of the caustic,
also including that of the peculiar motion of 
the foreground galaxy relative to the source and the observer,
on the disk plane.
The total observed flux will then be approximated by
\begin{equation}
\label{fx}
  F(t) = F_0\int_{-1}^1 A(x-\xi) P(\xi)d\xi.
\end{equation}
Here, $F_0$ is a constant representing the total
disk flux outside the caustic ($x < \xi$);
i.e., $\int P(\xi)d\xi = 1$, and
the amplification factor is given by
\begin{equation}
\label{ax}
  A(x-\xi) = \left\{ \begin{array}{lcl}
               1+\frac{\displaystyle k}{\displaystyle\sqrt{x-\xi}} 
                                       & \mbox{for} & x > \xi \\
              1                        & \mbox{for} & x \leq \xi, \\
                           \end{array} \right. 
\end{equation}
where $k$ denotes the amplification factor inside the caustic
($x>\xi$) and depends on 
%complex distributions of the caustic network
the spatial distribution of lensing stars and 
the number distribution of lens masses.
Note that GKS set $A= 0$ outside the caustics and, hence,
$k$ is arbitrary, %(included in $F_0$), 
which grossly simplified the analysis.
Also note that inclination effects are included in the expression
of $F_0$ and $V_{\rm caus}$ (discussed later).

We assume that $P(\xi)$ is represented by a continuous, piecewise
linear function,
\begin{eqnarray}
 P(\xi) = \frac{\xi_j-\xi}{\xi_j-\xi_{j-1}} P_{j-1}
         + \frac{\xi-\xi_{j-1}}{\xi_j-\xi_{j-1}} P_{j} \\
      \quad{\rm for}\quad\xi_{j-1}\leq\xi\leq\xi_j. \nonumber
\end{eqnarray}
For simplicity, we take equal-distant mesh points
normalized by the disk radius,
\begin{eqnarray}
  \xi_j = -1+ \left(j-\frac 12\right)\Delta\xi 
        = \frac{2j-J-1}{J} \\
  \quad(j=1, 2, \cdots, J), \nonumber
\end{eqnarray}
with $\Delta\xi\equiv 2/J$,
and we take the total number of mesh points ($J$) to be equal to 
that of the observed time sequences ($t_i$, $i=1, 2, \cdots, J$).
If we set $F_i = F(t_i)$, equation (\ref{fx}) leads
\begin{eqnarray}
\label{fkp}
   \vec{F} = F_0{\bf K} \vec{P}
\quad{\rm or}\quad
   F_i = F_0\sum_{j=1}^JK_{ij}P_j \\ 
\quad (i = 1, 2, \cdots, J) \nonumber
\end{eqnarray}
with 
\begin{equation}
   \vec{F} = (F_1, F_2, \cdots, F_J)^{\rm T}, \quad
   \vec{P} = (P_1, P_2, \cdots, P_J)^{\rm T},
\end{equation}
and $\bf{K}$ being a $J \times J$ matrix given in Appendix.

\subsection{Regularization Technique}

It seems straightforward to derive the emissivity distribution,
$P(\xi)$ or $P_j$ ($j=1, 2, \cdots, J$), from equation (\ref{fkp})
by calculating the inverse matrix of ${\bf K}$,
\begin{equation}
\label{soln1}
    \vec{P} = {\bf K}^{-1}(\vec{F}/F_0).
\end{equation}
However, this does not work efficiently, 
since the observational data usually
contain measuring errors, which will be greatly amplified
when calculating the inverse matrix (GKS).

To resolve this issue, a regularization technique was proposed.
First, we express the observed flux as
\begin{equation}
\label{fkpd}
   \vec{F} = F_0{\bf K} \vec{P} + \vec{\delta}
\quad{\rm or}\quad
   F_i = F_0\sum_{j=1}^J K_{ij}P_j + \delta_i,
\end{equation}
with $\delta_i$ being the error in measuring the flux, $F_i$.
Next, 
as a measure to evaluate how the $P(\xi)$ profile is smooth
we introduce a badness function,
\begin{equation}
   L(P) = \int \left[\frac{d^2 P(\xi)}{d\xi^2}\right]^2d\xi.
\end{equation}
(Note that $L(P)=0$ if $P(\xi)$ is a linear function of $\xi$.)

Then, the problem is reduced to determining the functional 
form of $P(\xi)$ which gives a minimum value of
\begin{equation}
\label{min}
    [(\vec{F}/F_0) - {\bf K}\vec{P}]^2 + \lambda L(\vec{P}).
\end{equation}
Namely, there are two important factors to be considered:
fitting to the given light curves ($\vec{F}$, the first term) and 
smoothing the emissivity profile ($\vec{P}$, the second term), where
$L(\vec{P})$ is the matrix form of $L(P)$, 
\begin{equation}
   L(\vec{P}) = \sum_j \left[\frac{P_{j+1}-2P_j+P_{j-1}}
                               {(\Delta\xi)^2}\right]^2\Delta\xi.
\end{equation}
The controlling parameter, $\lambda$, is called
the smoothing parameter.
In the limit of vanishing $\lambda$, the solution, $P(\xi)$, is given
by equation (\ref{soln1}) which can precisely reproduce the observed
light curves but is not always very smooth,
especially in cases with large measuring errors (GKS). 
In the limit of very large $\lambda$, on the other hand,
the solution is smoothest, since it gives
a straight line on the [$\xi,P(\xi)$] plane,
but may not give an excellent fit to the observed flux variations.
We thus need to choose a moderate value of $\lambda$
which satisfies
\begin{equation}
\label{chi2}
  {\frac 1J}\sum_{i=1}^J
      \left[\frac{1}{\delta_i}\left(
                  {F_i}-F_0\sum_{j=1}^JK_{ij}P_j
                                               \right)\right]^2 = 1.
%   (F - {\bf K}P)^2 = \sigma^2,
\end{equation}
In other words, the final solutions should be
{\it as smooth as possible under the constraint that a
reproduced flux variation agrees with the observed
variation within the error bars.}

After some algebra, minimizing equation (\ref{min})
is equivalent to setting
\begin{equation}
\label{soln2}
  ({\bf K}^{\rm T}{\bf K} + \lambda {\bf H}) \vec{P}
          = {\bf K}^{\rm T}(\vec{F}/F_0),
\end{equation}
where $\bf{H}$ is a matrix (see Appendix of GKS). 
The final solution, $P(\xi)$, is obtained successively
by solving equation (\ref{soln2}) for a given $\lambda$
satisfying equation (\ref{chi2}).

These procedures are performed for a specific value of $k$,
say, $k=0.5$ (which gives a variation with an amplitude of
$\sim$ 0.5 mag for $Q(r) \propto 1/r$).
We repeat the same procedures with different values of $k$
and determine the $k$-value so that $L(P)$ reaches the
minimum, yielding the smoothest $P(\xi)$.

\subsection{Transformation from $P(\xi)$ to $Q(r)$}

As stated in Introduction, our goal is to determine
the emissivity distribution of the disk as a function of $r$
(not $\xi$), where $r$ is the radial distance from the center
of the accretion disk in the unit of $r_0$ ($r=1$ at the
outer rim).  Thus, we need to transform $P(\xi)$ to $Q(r)$.

%Note that we do not {\it a priori} know $t=0$,
%the time when the caustic just crosses the disk center.
%In our technique, the disk center can be determined
%to be the place where $P(\xi)$ reaches the maximum.
%Likewise, $t_0$ can be fixed accordingly.

Note that the dimension of $Q(r)$ is flux; i.e., the
energy emitted per unit time from a unit surface area.
Then, $P(\xi)$ can be expressed as an integral of $Q(r)$ as
\begin{equation}
\label{pxi}
  P(\xi) = \int_{-1}^1 Q(r) d\eta
         = 2\int_{|\xi|}^1 Q(r)\frac{rdr}{\sqrt{r^2-\xi^2}}.
\end{equation}

It is of great importance to note that
equation (\ref{pxi})
takes the form of Abel's integral; that is, 
the inverse transformation is straightforward
(see, e.g., Binney,  Tremaine 1987).
Since we have two independent sets of $P(\xi)$,
one at $\xi < 0$ and the other at $\xi > 0$,
we can separately obtain two sets of $Q(r)$, $Q^-(r)$ and $Q^+(r)$;
namely,
\begin{eqnarray}
\label{qr}
  Q^-(r)=\frac{1}{\pi}\int_{-\infty}^{-r} 
           \frac{dP(\xi)}{d\xi} \frac{d\xi}{\sqrt{\xi^2-r^2}}
\quad{\rm and}\quad \nonumber \\
  Q^+(r)=\frac{1}{\pi}\int_r^\infty  \left[
          -\frac{dP(\xi)}{d\xi}\right] \frac{d\xi}{\sqrt{\xi^2-r^2}}.
\end{eqnarray}
Note that quite generally $dP/d\xi > 0$ for $\xi < 0$ and
$dP/d\xi < 0$ for $\xi > 0.$

The next procedure is used to derive an expression for
$Q^\pm_k = Q^\pm(r_k)$
in terms of $P_j = P(\xi_j)$ ($j=1,2, \cdots, J$) 
with $r_k \equiv (2k-1)/J$ for $k=1, 2, \cdots, J/2$.
We have
\begin{eqnarray}
  Q^-(r_k)\!\!&=&\!\!\frac{1}{\pi}\sum_{j=1}^{(J/2)-k}
           \frac{P_{j+1}-P_j}{\sqrt{\xi_{j+(1/2)}^2-r_k^2}} \nonumber \\ 
          & &+ \frac{1}{\pi}\frac{P_1}{\sqrt{1-r_k^2}} 
\end{eqnarray}
and
\begin{eqnarray}
  Q^+(r_k)\!\!&=&\!\!\frac{1}{\pi}\sum_{j=(J/2)+k}^{J-1}
           \frac{P_j-P_{j+1}}{\sqrt{{\xi}_{j+(1/2)}^2-r_k^2}} \nonumber \\
          & &+ \frac{1}{\pi}\frac{P_J}{\sqrt{1-r_k^2}} 
\end{eqnarray}
with ${\xi}_{j+(1/2)} \equiv |(\xi_j+\xi_{j+1})/2| = (2j-J)/J$,
and we used ${\xi}_{1/2} = -1$ and ${\xi}_{J+(1/2)}= +1$.
We can successively find $Q^\pm(r_k)$'s
for $k=1,2,\cdots, (J/2)-1$ from $P(\xi_j)$'s ($j=1,2,\cdots,J$).

\subsection{Producing Light Curves}

To test how the above procedures work,
we calculate the expected flux variations based on 
specific models for $Q(r)$
[hereafter denoted as $Q_{\rm model}(r)$ to distinguish
from the reconstructed values, $Q^\pm(r)$].
It is important to note that
accretion disk structure
can well be described as being self-similar (or
more precisely, each physical quantity is expressed
as a power-law function of the radius, see, e.g., 
Shakura, Sunyaev, 1973; Narayan, Yi 1995).  %; Kato et al. 1998).
We, hence, prescribe the emissivity distribution as
\begin{equation}
    Q_{\rm model}(r) \propto r^{-a}  \quad{\rm for~} 
		0.01 \leq r \leq 1.0
\end{equation}
with $a$ being a positive constant.
Note that $a=3$ for the standard-type disks,
whereas $Q_{\rm model}(r)$ 
is much flatter, $a \lsim 1$, in optically-thin, advection-dominated disks
(Manmoto et al. 1997).

Light curves are calculated by
\begin{eqnarray}
\label{ft}
  F(t_i)\!\!&=&\!\!F_0\int_0^1 rdr\int_0^{2\pi} d\theta
             A(x-r\cos\theta) Q_{\rm model}(r)  \nonumber \\ 
& & + \delta_i
% F(t_i)=F_0\int_{-1}^1d\xi\int_{-\sqrt{1-\xi^2}}^{\sqrt{1-\xi^2}}d\eta
%                  A(x-\xi) Q_{\rm model}(\sqrt{\xi^2+\eta^2}) + \delta_i
\end{eqnarray}
with the amplification factor [$A(x-r\cos\theta)$] being given by
equation (\ref{ax}).
We continuously change the
angular separation between the caustic and the center of the 
accretion disk (see Eq.[\ref{vcaus}]).  
The unit of time ($t_0$) is taken to be
the crossing time over which the caustic moves on 
the quasar disk plane from $\xi = 0$ to $+1$.
That is, $t_0 = r_0/V_{\rm caus}$, and, hence, 
$x(t) = t/t_0$ from equation (\ref{vcaus}).
% with $v_t$ being the transverse velocity
%of the caustic on the disk (source) plane.  
%The inclination effects can be included in the expressions
%for $F_0$ and $V_{\rm caus}$ (discussed later).
On the intrinsic light curve,
we add measuring errors, $\delta_i$.
We calculate three models for each $Q_{\rm model}(r)$ prescription;
$F(t)$ at each time
is randomly fluctuated around the mean value in a Gaussian way 
with assumed standard deviations, $\Delta m$ (in magnitude).

We plot in figure 2 how different microlens light curves are
produced by changes of the $Q_{\rm model}(r)$ prescription.
%%%%%%%%%% Figure 2 %%%%%%%%%%
\begin{figure}[htbp]
 \epsfxsize\columnwidth \epsfbox{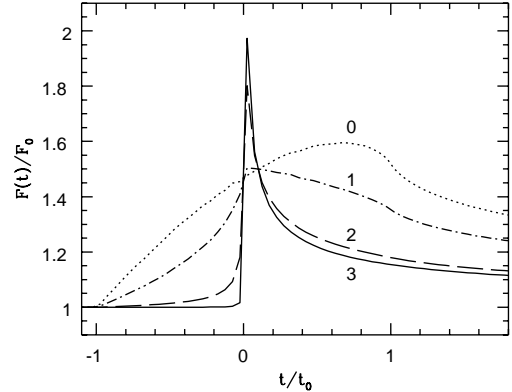}
\caption{Microlens light curves of disks whose
emissivity profiles are $Q(r) \propto r^{-a}$ with
$a=0, 1, 2$, and 3.
The values of $a$ are indicated in the figure.
The larger is $a$, the more sharply peaked is the light curve.
Note that the time of the peak flux is 
$t = 0$ when $a \geq 1$, but it is shifted to
$t \sim 0.7t_0$ when $a=0$.}
\end{figure}
%%%%%%%%%%%%%%%%%%%%%%%%%%%%%%
The normalizations are taken 
so as to give $F=1.5F_0$ at $t=0.1t_0$.
%in such a way that 
%the mean value of $F(t)$ averaged over a time span of $0.1t_0$ around 
%the time of the maximum flux.
%%%%%%%%%% Figure 3 %%%%%%%%%%
\begin{figure*}[t]
\begin{center}
 \epsfxsize\columnwidth \epsfbox{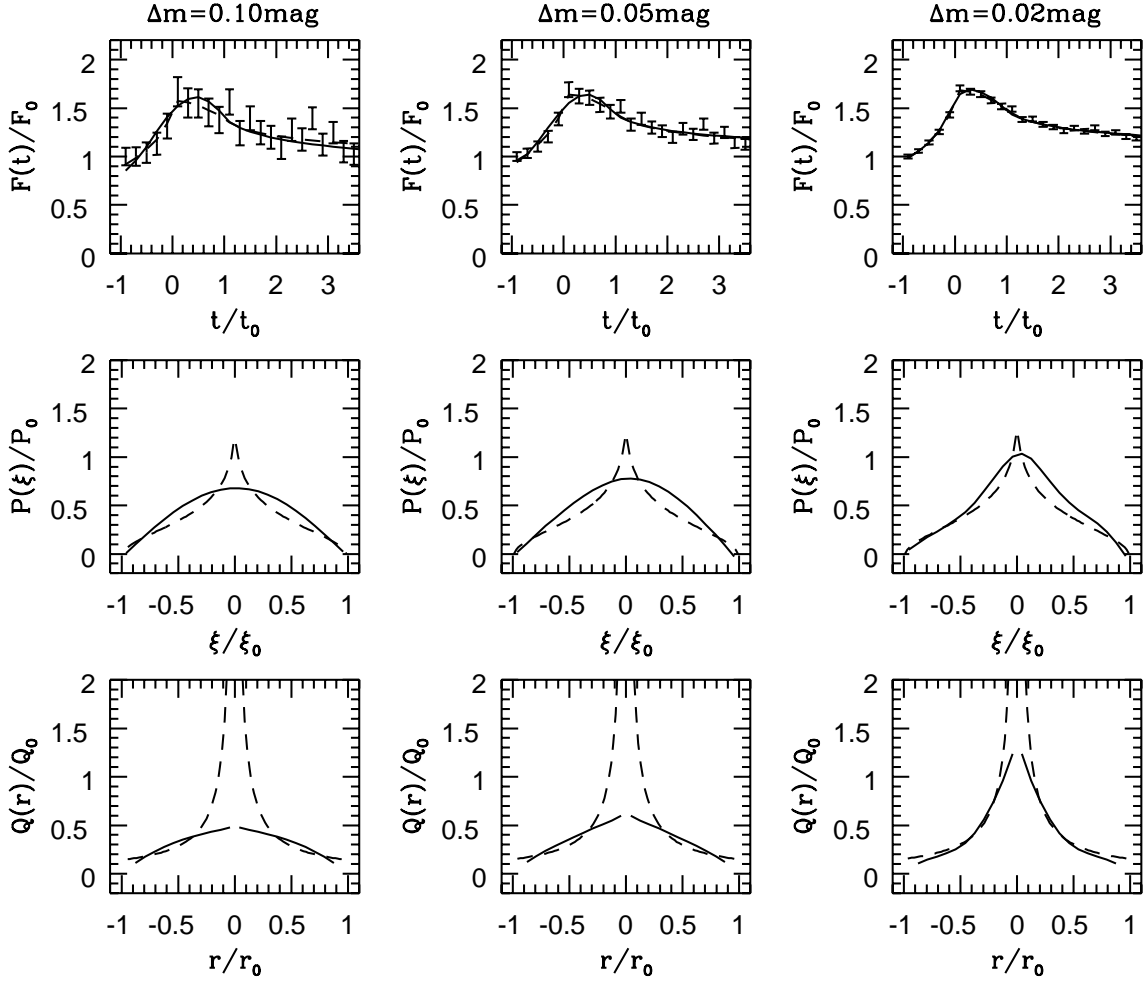}
\end{center}
\caption{Results of the reconstruction %with the original
for the cases with $Q_{\rm model} \propto 1/r$.
The dashed curves in the lower three panels are
the prescribed emissivity distribution ($Q_{\rm model}\propto 1/r$).
The solid curves in the lower panels at $r < 0$ (or $r > 0$) represent
the reconstructed $Q^-$ ($Q^+$) profiles
calculated from $P(\xi)$ at $\xi < 0$ ($\xi > 0$)
for three cases with $\Delta m=0.1$, 0.05, and 0.02 from the
left to the right, respectively.
The middle panels are the prescribed $P(\xi)$
calculated from $Q_{\rm model}(r)$,
and the reconstructed ones from the light curves above.
The upper three panels show the original light curves
calculated by the prescribed $Q_{\rm model}(r)$ (by the dashed lines)
with error bars,
and the reconstructed ones (by the solid lines).}
\end{figure*}
%%%%%%%%%%%%%%%%%%%%%%%%%%%%%%
Importantly, the peak luminosities are reached at the time of caustic
crossing over the disk center ($t=0$) only for $a\geq 1$,
while a disk with a flat emissivity profile (with $a=0$)
yields the peak at a later time, $t\sim 0.7t_0$.  This is
because for the flat disk the amplified area (which reaches its maximum
at $t=t_0$) is also an important factor.  For $a\geq 1$, in contrast,
the emergent flux is rather insensitive to the extent of the amplified area,
since a large fraction of radiation 
comes from the very center of the disk.

To summarize, those cases with $a=0$ and $a\geq 1$ give distinct
flux variations.
It is thus important to distinguish these two critical cases
in disk mapping.

\section{Results of Image Reconstruction}

The results of the reconstruction are displayed in figure 3
for the case with $Q_{\rm model}(r) \propto 1/r$.
The prescribed emissivity profile is illustrated by the dashed lines
in the lower three panels.
To calculate the flux variation, we set
a constant time interval, $\Delta t= 0.2t_0$, and the total number
of observations is $J=24$.
The calculated (i.e., `observed') flux at each time
is displayed together with error bars
in the upper three panels by the dashed lines
for different magnitudes of the mean errors,
$\Delta m$ = 0.1 (left), 0.05 (middle), and 0.02 (right), respectively.  
The dashed lines in the middle ones 
show the one-dimensional emissivity, $P(\xi)$,
calculated from $Q_{\rm model}(r)$.

Similarly,
the reconstructed $F(t)$ and $P(\xi)$
are shown in the upper and middle panels with the solid lines.
The two solid lines in the lower panels display
$Q^-(r)$ at $r < 0$ and $Q^+(r)$ at $r > 0$, respectively.
Note that the reconstructed $k$-value
does not always reproduce the original value of $k=0.50$;
we find $k=0.61$, 0.51, and 0.47 for $\Delta m = 0.10$, 0.05, and 0.02,
respectively.
The reconstructed flux variations and emissivity profiles are
not very sensitive to these small deviations in the $k$-values,
however.

For $a=1$, large errors ($\Delta m=0.1$ mag)
tend to produce a rather flat $P(\xi)$ profile and
thus a smooth $Q(r)$.  This is because
a flat model [$P(\xi) =$ const.] is compatible with
the light curves within the error bars so that the technique
prefers a flatter $P(\xi)$ profile (see figure 2).  
Note again that the
peak shifts from $t=0$ in the original light curve to 
$t\sim 0.7 t_0$ in the reconstructed one.
Such a problem does not arise for those cases with
smaller errors.  To reproduce a steep $Q(r)$ profile up to
the inner parts, $\Delta m = 0.02$ mag is necessary.
It depends on the observing intervals how close to the origin
the mapping technique can reproduce the original image.  
Surely, frequent observations, especially at around the peak flux,
are preferable (GKS).

Figures 4 and 5 display the results of those cases with
$Q_{\rm model}\propto 1/r^2$, and $\propto 1/r^3$, respectively.  
In each figure,
we omit the cases with $\Delta m$ = 0.05 and $P(\xi)$ plots
to avoid any complications.
%%%%%%%%%% Figure 4 %%%%%%%%%%
\begin{figure}[htbp]
  \epsfxsize\columnwidth \epsfbox{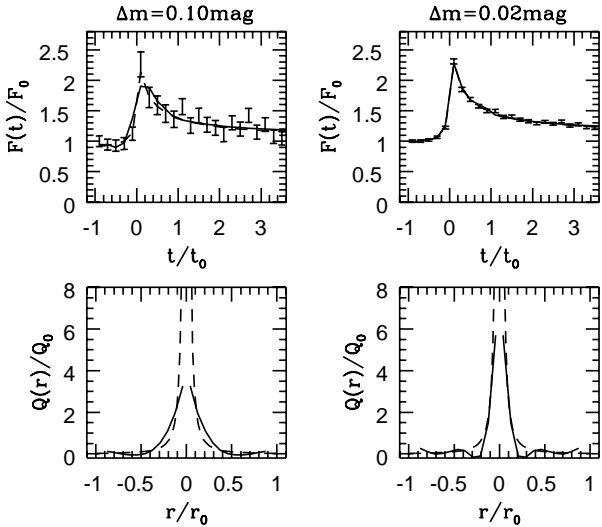}
\caption{Results of the reconstruction and the original
for the cases with $Q_{\rm model} \propto 1/r^2$.
The lower two panels are
the prescribed emissivity distribution (by the dashed lines)
and the reconstructed one (by the solid lines)
for two cases with $\Delta m=0.1$ (left) and 0.02 (right),
respectively,.
The upper two panels show the original light curves
(by the dashed lines) with error bars,
and the reconstructed ones (by the solid lines).}
\end{figure}
%%%%%%%%%%%%%%%%%%%%%%%%%%%%%%
%%%%%%%%%% Figure 5 %%%%%%%%%%
\begin{figure}[t]
 \epsfxsize\columnwidth \epsfbox{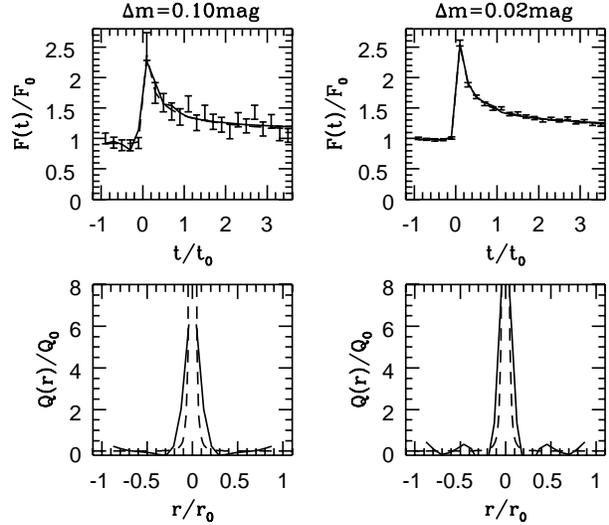}
\caption{Same as figure 4, but 
for cases with $Q_{\rm model} \propto 1/r^3$.}
\end{figure}
%%%%%%%%%%%%%%%%%%%%%%%%%%%%%%

The reconstructed $k$-values are
0.63 ($\Delta m = 0.10$) and 0.74 ($\Delta m=0.02$) for $a=2$, and
0.76 ($\Delta m = 0.10$) and 0.78 ($\Delta m = 0.02$) for $a=3$, 
respectively.  
These all deviate greatly from the original value of $k=0.5$,
particularly when $a$ is large or $\Delta m$ is small.
Nevertheless,
the inner steep rise parts are reasonably well reproduced by
the mapping.  The wing parts are poorly reproduced
for a steep $Q(r)$ profile.  In fact,
$Q(r)$ sometimes goes below zero.
This tendency is rather enhanced for small $\Delta m$'s.

To conclude,
in cases with sharply peaked emissivity profiles
we can still obtain reliable information regarding
the extent of the substantially emitting region,
although we cannot trust the results about the outer zones
surrounding the central bright zone.

\section{Summary and Discussion}

Let us consider specifically the case of Einstein Cross and thus
insert the model parameters relevant to this source.
The Einstein-ring radius on the source plane is
\begin{eqnarray}
 r_{\rm E} \equiv \theta_{\rm E}D_{\rm os} 
    \!\!&=&\!\! \left[\left(\frac{4GM_{\rm lens}}{c^2}\right)
            \left(\frac{D_{\rm ls}D_{\rm os}}
                       {D_{\rm ol}}\right)\right]^{1/2} \nonumber \\
    \!\!&\sim&\!\!1.5\times 10^{17}
            \left(\frac{M_{\rm lens}}{M_\odot}\right)^{1/2}{\rm cm},
\end{eqnarray}
where
$M_{\rm lens}$ is the typical mass of a lens star, 
and $D_{\rm ls}$, $D_{\rm os}$, and $D_{\rm ol}$ 
represent the angular diameter distances from lens to source, 
from observer to source, and from observer to lens, respectively.
%(see Paczy\'nski 1986)
To evaluate these distances, we assume
the redshifts corresponding to the distances from the observer to the quasar
and from the observer to the lens of, %and from the lens to the quasar of,
$z_{\rm os} = 1.675$ and $z_{\rm ol} = 0.039$, %and $z_{\rm ls} = 1.575$, 
respectively (see Irwin et al. 1989),
and also assumed an Einstein-de Sitter universe and
Hubble's constant to be $H_{\rm 0} \sim 60 {\rm km~s^{-1} Mpc^{-1}}$, 
according to Kundi\'c et al. (1997).

Another important length is $r_{\rm cross}$,
the caustic crossing length over the quasar image plane per observational
time interval, $\Delta t$;
\begin{eqnarray}
\label{cross}
 r_{\rm cross}\!\!&=&\!\!
   v_{\rm t}\Delta t \frac{D_{\rm os}}{D_{\rm ol}} \nonumber \\
      \!\!&\sim&\!\!2.0\times 10^{13}
              \left(\frac{v_{\rm t}}{300~{\rm km~s}^{-1}}\right)
              \left(\frac{\Delta t}{1~{\rm d}}\right) {\rm cm},
%      \sim 4.1\times 10^{13}
%             \left(\frac{v_{\rm t}}{600~{\rm km~s}^{-1}}\right)
%             \left(\frac{\Delta t}{1~{\rm d}}\right) {\rm cm},
\end{eqnarray}
where $v_{\rm t}$ is the transverse velocity of 
the lens on the lens plane 
($V_{\rm caus} \equiv v_{\rm t}D_{\rm os}/D_{\rm ol}$).
%including the transverse velocity of the peculiar motion of 
%the foreground galaxy relative to the source and the observer.
Surprisingly, this is comparable to the Schwarzschild radius,
$r_{\rm g}\simeq 3\times 10^{13}(M_8)^{-1}$cm ($\sim 2$AU)
for a $10^8M_8$ black hole and is much smaller than $r_{\rm E}$.
% \begin{equation}
%  r_{\rm cross} \sim 20AU \simeq 10 r_{\rm g}  M_8^{-1},
% \end{equation}
%for $v_{\rm t} \sim$ 600~km~s$^{-1}$ and $\Delta t \sim 7$d.
Thus, by weekly observations can one determine 
the disk emissivity distributions
on length scales of $\sim$ 10AU or
$\sim 5 r_{\rm g}(M_8)^{-1}$ for
$v_{\rm t} \sim$ 300~km~s$^{-1}$.  %(Yonehara et al. 1998).

To summarize, we have improved the reconstruction technique
previously developed by GKS in such a way that
a direct comparison with accretion disk models is possible.
We have found that for deriving the emissivity distribution as a function
of $r$ on scales down to several to ten AUs, we need an accuracy of
$\Delta m \lsim 0.02$mag and a sampling interval within one week.
In cases in which the emissivity profile is rather centrally
peaked (i.e., if $a > 1$),
we can still reproduce the inner bright zone reasonably well,
but cannot trust the results of the outer zone.

It might be noted that the problem treated here seems to be closely
related to that of reconstructing stellar brightness profile
(e.g., limb darkening) from photometry of Galactic
microlensing events (see Gaudi, Gould, 1999, and references therein).
The present technique may be used to analyze such data.

The inclination of the disk has two important effects.
If the disk plane is tilted by an angle of $i$
with respect to the line of sight,
the apparent disk flux will be reduced by a factor of $(\cos i)^{-1}$.
This affects the normalization constant, $F_0$.
Further, if the
disk is tilted by an angle of $i_\parallel$
with respect to the direction of the motion of a caustic,
the apparent transverse velocity of the caustic will be increased
by a factor of $(\cos i_\parallel)^{-1}$.
This affects the caustic crossing, $V_{\rm caus}$,
and thus the time and length scales, $t_0$ and $r_0$.
In other words, the normalization constants both in the
ordinate and abscissa in the $(t, F)$ diagram and thus 
in the $(r, Q)$ diagram are subject to the inclination angles.
There remain ambiguities in the normalizations
of $r_0$ and $Q_0$.  Since it is difficult to evaluate these
inclination angles and other uncertain factors,
we rather focused our effort on the shape of 
the non-dimensional emissivity distribution, $Q/Q_0$,
as a function of $r/r_0$.

%%%%%%%%%% Figure 6 %%%%%%%%%%
\begin{figure}[htbp]
 \epsfxsize\columnwidth \epsfbox{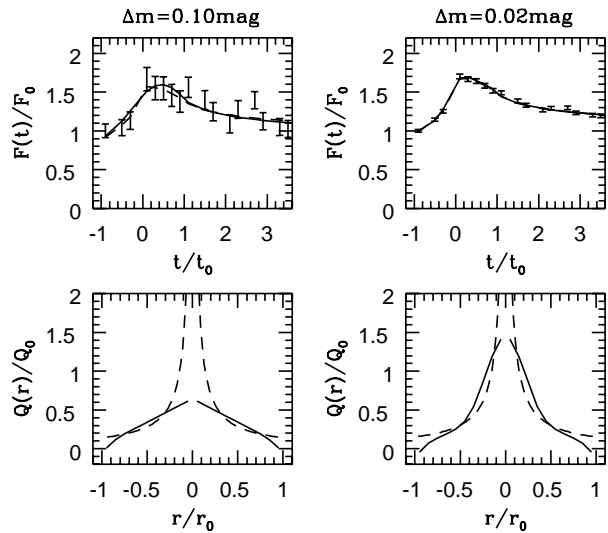}
\caption{Same as figure 4, but
for the unevenly sampled case.  
The given profile is $Q_{\rm model} \propto 1/r$.}
\end{figure}
%%%%%%%%%%%%%%%%%%%%%%%%%%%%%%

We must consider the fact that
real photometric light curves will be unevenly sampled
due to gaps in the observing schedule caused by
various factors, including weather.  Therefore,
an even more realistic case could be made by making a model
of such gaps.  To see such an effect,
we made a light curve in which a certain fraction, $\sim 25$\%,
of the regularly spaced light curve samples is removed stochastically.
The results for the cases with $Q_{\rm model} \propto 1/r$
are displayed in figure 6.
The reconstructed $k$-values are
0.48 ($\Delta m = 0.10$) and 0.415 ($\Delta m=0.02$), respectively.
Because there are no significant changes,
as long as frequent observations are made around the peak,
a lack of data (uneven sampling) should not cause any serious problem.
Obviously, however, 
poor sampling rates lead to a poor spatial resolution in the disk mapping
(see equation ~\ref{cross}).
The results would sensitively depend on whether or not there are
enough observation runs at the times around the peak.

 HST and AXAF observations of the Einstein Cross are scheduled.
If observed with multi-wavelength bands,
a microlensing event
should first clearly resolve the multi-wavelength radiation properties of
a disk in a distant quasar on length scales down to several AUs.

\par
\vspace{1pc} \par
The authors would like to express their thanks to 
Joachim Wambsgan{\ss} for valuable suggestions
and an anonymous referee for useful comments,
which helped in making the revised version.
This work was supported in part by the Grants-in Aid of the
Ministry of Education, Science, and Culture of Japan
(10640228, SM) and
by Research Fellowships of the Japan Society for the
Promotion of Science for Young Scientists, 9852 (AY).

\section*{Appendix}

Equation (\ref{fx}) can be rewritten as
\begin{equation}
  F_i = \sum_{j=1}^J f_{ij},
\end{equation}
where $f_{ij}$ is non-zero only for $x_i > \xi_{j-1}$
and is
\begin{eqnarray}
  f_{ij} 
  \!\!&=&\!\!\int_{\xi_{j-1}}^{\min(\xi_j,x_i)}A(x-\xi)P(\xi)d\xi \nonumber \\
         \!\!&=&\!\!\int_{\xi_{j-1}}^{\min(\xi_j,x_i)}
            \left(1+\frac{k}{\sqrt{x_i-\xi}}\right) \nonumber \\
            & & \times \left(\frac{\xi_j-\xi}{\Delta\xi} P_{j-1}
                + \frac{\xi-\xi_{j-1}}{\Delta\xi} P_{j}\right)d\xi,
\end{eqnarray}
with $\Delta\xi\equiv \xi_{j}-\xi_{j-1}$.
We introduce matrices $E_{ij}$ and $D_{ij}$ such that
\begin{equation}
\label{ed}
  f_{ij} = E_{ij}P_{j-1} + D_{ij}P_j.
\end{equation}
We then have
(for $x_i > \xi_{j-1}$)
\begin{eqnarray}
\label{ee}
  E_{ij}\!\!&=&\!\!\frac{1}{\Delta\xi} \int_{\xi_{j-1}}^{\min(\xi_j,x_i)}
     \left(\xi_j-\xi+k\frac{\xi_j-\xi}{\sqrt{x_i-\xi}}\right)d\xi \nonumber \\
         \!\!&=&\!\!\frac{1}{\Delta\xi}\int_{\Delta'\xi}^{\Delta\xi}
           \left(s+k\frac{s}{\sqrt{x_i+s-\xi_j}}\right)ds \nonumber \\
         \!\!&=&\!\!\frac{1}{\Delta\xi}
           \left[\frac{s^2}{2}\right.+2ks\sqrt{x_i+s-\xi_j} \nonumber \\
         & & ~ ~ ~ ~ ~ ~ ~ ~ ~ ~ -\left.\frac{4k}{3}(x_i+s-\xi_j)^{3/2}
                                \right]_{\Delta'\xi}^{\Delta\xi} \nonumber \\
         \!\!&=&\!\!\frac{(\Delta\xi)^2-[\rho(\xi_j-x_i)]^2}{2\Delta\xi}
                + 2k {\sqrt{x_i-\xi_{j-1}}} \nonumber \\
         & & -\frac{4k}{3} \frac{(x_i-\xi_{j-1})^{3/2}-[\rho(x_i-\xi_j)]^{3/2}}
                      {\Delta\xi},
\end{eqnarray}
with 
$s\equiv \xi_j-\xi$, $\Delta'\xi \equiv \rho(\xi_j-x_i)$ and
\begin{equation}
\rho(x_i-\xi_j) \equiv \left\{ \begin{array}{lcl}
                     x_i-\xi_{j} & \mbox{for} & x_i > \xi_j \\
                         0       & \mbox{for} & x_i \leq \xi_j. \\
                           \end{array} \right.
\end{equation}
Similarly, we find
(for $x_i > \xi_{j-1}$)
\begin{eqnarray}
\label{dd}
  D_{ij}\!\!&=&\!\!\frac{1}{\Delta\xi} \int_{\xi_{j-1}}^{\min(\xi_j,x_i)}
           \left(\xi-\xi_{j-1}
                +k\frac{\xi-\xi_{j-1}}{\sqrt{x_i-\xi}}\right)d\xi \nonumber \\
         \!\!&=&\!\!\frac{1}{\Delta\xi}\int_0^{\Delta''\xi}
           \left(u+k\frac{u}{\sqrt{x_i-u-\xi_{j-1}}}\right)du \nonumber \\
         \!\!&=&\!\!\frac{1}{\Delta\xi}
           \left[\frac{u^2}{2}\right.-2ku\sqrt{x_i-u-\xi_{j-1}} \nonumber \\
         & & ~ ~ ~ ~ ~ ~ ~ ~ ~ ~ -\left.\frac{4k}{3}(x_i-u-\xi_{j-1})^{3/2}
                                           \right]_0^{\Delta''\xi} \nonumber\\
         \!\!&=&\!\!\frac{(\Delta''\xi)^2}{2\Delta\xi}
          -2k\frac{\Delta''\xi}{\Delta\xi} \sqrt{\rho(x_i-\xi_j)} \nonumber \\
         & & +\frac{4k}{3} \frac{(x_i-\xi_{j-1})^{3/2}-[\rho(x_i-\xi_j)]^{3/2}}
                      {\Delta\xi},
\end{eqnarray}
with $u\equiv \xi-\xi_{j-1}$
and $\Delta''\xi \equiv\min(\Delta\xi, x_i-\xi_{j-1})$.
For $x_i \leq \xi_{j-1}$ both are zero, $E_{ij} = D_{ij} = 0$.
%If we set %$E_{i,J+1} = \Delta\xi/2$ and 
%$P_0=0$, 
Since we have from equation (\ref{ed})
\begin{equation}
  f_{ij} = \left(E_{i,j+1} + D_{ij}\right)P_j \equiv K_{ij}P_j,
\end{equation}
for $(i,j)=(1,1), (1,2), \cdots, (J,J)$, 
it is straightforward to derive expression for $\bf{K}$,
\begin{equation}
K_{ij} = E_{i,j+1} + D_{ij},
\end{equation}
by inserting equations (\ref{ee}) and (\ref{dd}).

\section*{References}
\small

\re
 Binney J., Tremaine S. 1987, Galactic Dynamics (Princeton Univ. Press,
 Princeton) p651

\re
 Blandford R.D., Hogg D.W. 1995, in IAU Sympo. 173,
  Astrophysical Application of Gravitational Lensing,
  ed C.S. Kochanek, J.N. Hewitt
  (Kluwer, Dordrecht) p355

\re
 Chang K., Refsdal S. 1979, Nature 282 561

\re
 Chang K., Refsdal S. 1984, A\&A 132, 168

\re
 Corrigan R.T. et al. 1991, AJ 102, 34

\re
 Gaudi B.S., Gould A. 1999, ApJ 513, 619

\re
 Grieger B., Kayser R., Refsdal S. 1988, A\&A 194, 54

\re
 Grieger B., Kayser R., Schramm T. 1991, A\&A 252, 508 (GKS)

\re
 Horne K., 1985, MNRAS 213, 129

\re
 Houde M., Racine R. 1994, AJ 107, 466

\re
 Huchra J. et al. 1985, AJ 90, 691

%Ichimaru S. 1977, ApJ 214, 840

\re
 Irwin M.J., Webster R.L., Hewett P.C., Corrigan R.T.,  
Jedrzejewski R.I. 1989, AJ 98, 1989

\re
 Jaroszy\'nski M., Wambsganss J.,  Paczy\'nski B. 1992,
ApJ 396, 65L

\re
 Kundi\'c T. et al. 1997, ApJ 482, 75 

\re
 Manmoto T., Mineshige S., Kusunose M., 1997, ApJ 489, 791

\re
 McKernan B., Yaqoob T. 1998, ApJ 501, L29

\re
Narayan R., Yi I. 1995, ApJ 452, 710

\re
 Ostensen R., et al. 1996, A\&A 309, 59 

\re
 Rauch K.P., Blandford R.D. 1991, ApJ, 381, L39

\re
 Shakura N.I., Sunyaev R.A. 1973, A\&A 24, 337

\re
 Wambsganss J., Paczy\'nski B. 1991, AJ 102, 864

\re
 Wambsganss J., Paczy\'nski B. 1994, AJ 108, 1156

\re
 Yonehara A., Mineshige S., Fukue J., Umemura M., Turner E.L. 
 1999, A\&A 343, 41

\re
 Yonehara A., Mineshige S., Manmoto T., Fukue J., Umemura M., Turner E.L. 
1998, ApJ 501, L41; Erratum 511, L65

\label{last}

\end{document}